\documentclass[twocolumn,aps,prl,showpacs,superscriptaddress,amsfonts,amssymb,amsmath]{revtex4}
\usepackage{graphicx}
\begin{document}
\title{Do intradot electron-electron interactions induce dephasing?}
\author{Zhao-tan Jiang$^1$, Qing-feng Sun$^{1,\ast}$, X. C. Xie$^{2,3}$, and Yupeng
Wang$^{1,3}$}
\affiliation{Beijing National Lab for Condensed Matter Physics and
Institute of Physics, Chinese Academy of Sciences, Beijing
100080, China \\
$^2$Department of Physics, Oklahoma State University, Stillwater,
Oklahoma 74078 \\
$^3$International Center for Quantum structures, Chinese Academy
of Sciences, Beijing 100080, China }
\date{\today}

\begin{abstract}
We investigate the degree of coherence of electronic transport
through a quantum dot (QD) in the presence of an intradot
electron-electron interaction. By using an open multi-terminal
Aharonov-Bohm (AB) setup, we find that the intradot interaction
does not induce any dephasing effect and the electron transport
through the QD is fully coherent. We also observe that the
asymmetric amplitude of the AB oscillation in the conductance
through the two-terminal AB setup originates from the interplay
between the confined structure and the electron-electron
interaction. Thus, one can not associate a dephasing process with
this asymmetric amplitude, as has been done in previous studies.
\end{abstract}


\maketitle

How electron-electron (e-e) interactions influence the phase
coherence of electronic transport through mesoscopic systems, e.g.
quantum dots (QD), has been one of the most significant and
challenging issues from fundamental physics point of view as well
as for realization of quantum devices. By embedding a QD in one
arm of an Aharonov-Bohm (AB) interferometer, it has been
experimentally demonstrated that the transport through the QD is
at least partially coherent despite the existence of a strong
intradot e-e interaction\cite{n1,n2,n3}. Naturally, one may ask
further about the coherence rate, i.e. how much coherence is
maintained in such a tunneling process? Recent theoretical
studies\cite{n4,n5,n6} addressed this question and arrived at the
conclusion that an intradot e-e interaction will induce partial
dephasing. According to these studies, the simple reason for the
partial incoherence is from the spin-flip process.\cite{n4,n5,n6}
In this process, for instance, a spin-up electron enters the QD
and a spin-down electron exits, whereas the spin in the QD gets
flipped. In other words, the traversing electron has left a trace
in the QD (i.e. the ``environment''), causing dephasing. However,
this simple intuitive spin-flip picture is not quite transparent,
as mentioned by the same authors\cite{n4,n5,n6}. For instance, one
possible drawback in the argument given above is that only one electron in
the leads is involved, i.e. neglecting of many-body features in the
leads. Another shortcoming is that one has to artificially divide
the successive tunneling process into a series of second-order
processes. The tunneling process is a continuous one and it is not
clear that such a division is a proper procedure.
In order to further clarify the question of the dephase by
the e-e interaction, Konig and Gefen\cite{n5} have analyzed the
transport behavior of the two-terminal AB interferometer embedded
with a QD. They found that the amplitude of AB oscillations in the
conductance is suppressed compared with the non-interacting case.
Furthermore, they predicted that, as a consequence of dephasing,
an asymmetry would appear around resonant peaks. This prediction
has been observed in a recent experiment,\cite{n8} lending a
strong support to the belief of interaction induced dephasing.

However, we question whether the two-terminal AB setup is a proper
geometry to study the dephasing effect. Since in this confined and
closed structure, the repeated reflection and tunneling processes
are plentiful. Is it possible that it is this confinement, not the
dephasing that suppresses the amplitude of AB oscillations and
makes them asymmetric? It is known that, due to the confinement,
the two-terminal AB setup has the phase locking
effect,\cite{n1,n2,n9,n10} and the phase of the transmission
amplitude can not be determined. In this Letter, for the first
time, we investigate the coherence issue in an open multi-terminal
AB setup. We find that the intradot e-e interaction does not
induce any dephasing effect and the electronic transport through
the QD maintains fully coherence. To a larger extent, our results
will shed light on the issue of low temperature saturation of
phase coherence time\cite{webb} observed in many solid-state
samples.\cite{add2note}

We consider an open multi-terminal AB setup (see Fig.1), mimic
experimental configurations\cite{n1,n2,n3,n8}.
Four extra leads are attached to four sites and a QD
is embedded in the lower arm. In general, the extra leads may
introduce dephasing effects into the system as demonstrated by
Buttiker.\cite{n11} In order to avoid the
dephasing by the extra leads, we keep their
voltages to be the same as that for the drain (i.e. $V_i =V_d\equiv 0$, with
$i=1$, 2, 3, and 4). Electrons will only inject from the
source into the AB ring, and leave from the drain and the other four extra
leads. In this way, due to the current bypass effect,
the processes of repeated reflection and tunneling are strongly
suppressed, and the first-order tunneling process dominates.

In order to quantitatively describe the phase coherence, we
introduce the coherence rate parameter $r_T(\epsilon)$:
$r_T(\epsilon) \equiv \frac{T_1(\epsilon)}{2\sqrt{T_{ref}
T_d(\epsilon)}}$, where $T_1(\epsilon)$ is the first-order AB
oscillation amplitude of the transmission probability
$T(\epsilon)$ from source to drain, $T_{ref} =|t_{ref}|^2
$ and $T_d$ are transmission probabilities through the reference arm
and the QD, respectively. $T_d= T_{co} +T_{in}$, where
$T_{co}=|t_{co}|^2$ and $T_{in}$ are the coherent and incoherent
components of the transmission probability through the QD. If to
assume only the first-order tunneling process exist, then
$T_1=2|t_{ref} t_{co}|$ and $r_T$ reduces to:
\begin{eqnarray}
r_{T} = \frac{|t_{co}|}{\sqrt{|t_{co}|^2+T_{in}}}.
\end{eqnarray}
Obviously, in this case the value of $ r_T$ directly reflects the
degree of coherence. If $r_T =1$, it is fully coherent; on the
other hand, if $r_T=0$, it is completely incoherent. Of course,
when the higher-order tunneling processes are not
negligible, Eq.(1) is no longer valid. In this case, $r_T$ does
not reflect the degree of coherence. In this work we design a system
(i.e. open multi-terminal AB setup) in which the first-order
tunneling process dominates, and we carry out a study of the coherence
rate $r_T$ in such a system.

The entire system of the AB interferometer considered here (see
Fig.1) is modelled by the Hamiltonian
\begin{eqnarray}
H=H_{0}+H_{R}+H_{T}. \label{eq:1}
\end{eqnarray}
Here $H_{0}=\sum_{\alpha ,k ,\sigma}{\epsilon_{\alpha k}
c^+_{\alpha k \sigma}c_{\alpha k \sigma}}$ describes the
noninteracting electrons in the source, drain, and four extra
leads with $\alpha$=$s$, $d$, and 1,$\cdots$,4, respectively. The
Hamiltonian of the isolated AB ring, composed of the QD and four
sites, is represented by: $
H_{R}=\sum_{\beta,\sigma}{\epsilon_{\beta\sigma}d^+_{\beta\sigma}d_{\beta\sigma}}
+Un_{{d\uparrow}}{n}_{d{\downarrow}}
+\sum_{\sigma}\left(t_{r}e^{i\phi}d_{2\sigma}^+d_{3\sigma}
+t_{1}d_{1\sigma}^+d_{d\sigma}+t_{4}d_{4\sigma}^+d_{d\sigma}+H.c.\right)
$, where $d_{\beta \sigma}^+$ ($d_{\beta \sigma}$) are the
creation (annihilation) operators in the four sites
($\beta=1,...,4$), or in the QD ($\beta=d$).\cite{addnote} The QD
includes a single energy level $\epsilon_{d\sigma}$ having spin
index $\sigma$ and an intradot e-e interaction $U$. To account for
the system threaded by a magnetic flux $\Phi$, a phase factor
$e^{i\phi}$ with $\phi =2\pi\Phi/\Phi_0$ is attached to the
hopping matrix element $t_r$ through the reference arm. The last
term in Eq.~(\ref{eq:1}), $ H_T=(\sum_{\alpha, \beta, k,
\sigma}t_{\alpha \beta }c_{\alpha k\sigma}^+d_{\beta
\sigma}+\sum_{\gamma, k,\sigma}t_{\gamma \gamma }c_{\gamma
k\sigma}^+d_{\gamma \sigma}+H.c.), $ describes the tunneling
between the AB ring and the leads. Here $t_{\alpha \beta }$
represents the coupling matrix elements between lead $\alpha$=$s$
($d$) and site $\beta$=$1,2$ ($3,4$), while $t_{\gamma \gamma }$
denotes that between the $\gamma$th extra lead and the $\gamma$th
site with $\gamma=1,...,4$.

By using the standard Keldysh non-equilibrium Green's function
method,\cite{book1} the conductance of the drain current $I_d$
versus the source voltage $V_s$ can be derived as:\cite{n13}
\begin{eqnarray}
\label{eq:a} {G}\equiv \frac{d I_d}{d V_s} =
\frac{-e}{\hbar}\sum_{n,m,\sigma} Im \int{\frac{d\epsilon}{2\pi}}
\Gamma_{nm}^d \frac{d}{d V_s} \left[{G}_{mn}^< + 2 f_d(\epsilon)
G_{mn}^r \right],
\end{eqnarray}
where the coupling strength $\Gamma_{nm}^d \equiv 2\pi \sum_k
t^*_{dn}t_{dm}\delta(\epsilon-\epsilon_{dk})$, and
$f_{d/s}(\epsilon)$
% $ =\left\{exp[(\epsilon-\mu_{d/s})/(k_B {\cal T})]+1\right\}^{-1}$
is the Fermi distribution function of the drain/source.
%with chemical potential $\mu_{d/s}$.
$G^r_{nm}(\epsilon)$ and $G^<_{nm}(\epsilon)$ are the standard
retarded and Keldysh Green's functions.\cite{n13} They are
5$\times$5 matrices and the index $n,m =1,...,4$ for the
corresponding sites and $n,m=5$ for the QD. We solve the Green's
functions by the following procedures. First, the isolated QD
Green's functions is exactly obtained: $g_{55}^r=
[\epsilon-\epsilon_{d\sigma}-U+Un_{\bar{\sigma}}]/
[(\epsilon-\epsilon_{d\sigma})(\epsilon-\epsilon_{d\sigma}-U)]$.
Second, using the Dyson equation $G^r =g^r +G^r \Sigma^r g^r$ and
the Keldysh equation $G^<=G^r\Sigma^<G^a$, the Green's functions
$G^r$ and $G^<$ of the whole system can be derived.\cite{note1} As
the last step, $n_{\sigma}$, the intradot electron occupation
number for spin state $\sigma$, is solved self-consistently with
the self-consistent equation $n_{\sigma} =-i\int
\frac{d\epsilon}{2\pi} G^<_{55}(\epsilon)$.

In order to study the degree of coherence of electronic transport
through an interacting QD, we numerically study the linear
conductance and the coherence rate. In the numerical calculations,
we choose a very weak $t_r=0.001$ and low temperature $k_B T
=0.01$. The four sites' energy levels are chosen to be
$\epsilon_1=-\epsilon_2=\epsilon_3=-\epsilon_4=2$.\cite{note2} We
also set $\Gamma_{11}^s=\Gamma_{22}^s=\Gamma_{33}^d =\Gamma_{44}^d
\equiv \Gamma^{s} \equiv \Gamma^{d}=10$ as the energy unit, and
$\Gamma_{11}=\Gamma_{22}=\Gamma_{33}=\Gamma_{44}\equiv \Gamma$.
Here $\Gamma_{\gamma\gamma}\equiv 2\pi \sum_k
|t_{\gamma\gamma}|^2\delta(\epsilon-\epsilon_{\gamma k})$ describe
the coupling strength between the extra leads and the
corresponding sites. A larger $\Gamma$ gives a stronger coupling,
and enhances the current bypass effect, thus, the first-order
tunneling process dominates. In the limit case of $\Gamma
\rightarrow \infty$, only first-order tunneling process survives.

We first investigate the spin-degenerate case with no magnetic
field in the QD. However, we are still under AB configuration with
non-zero magnetic flux $\Phi$ passing through the AB ring. The
total linear conductance $G$ versus the QD's level $\epsilon_d$
for the open AB setup exhibits two Coulomb oscillation peaks at
$\epsilon_d=0$ and $-U$ (see Fig.2a). At a fixed $\epsilon_d$, $G$
versus the magnetic flux $\phi$ shows periodic oscillations with a
period of $2\pi$ (see Fig.2b). Due to this periodic oscillations,
$G(\phi)$ can be expanded in a Fourier series: $G(\phi) = G_0 +
G_1 cos (\phi +\varphi_1) + G_2 cos (2\phi +\varphi_2) + ...$,
where $G_1$ is the first-order amplitude of AB oscillations. Since
$t_r$ is chosen as a small parameter, $G_1 \propto t_r$. When the
AB setup is decoupled with the reference (or the QD), i.e. when
$t_r=0$ (or $t_1=t_4=0$), the conductance $G_d$ (or $G_{ref}$)
through the QD (or the reference arm) can also be obtained. This
enable us to define an experimental measurable conductance
coherence rate $r_G(\epsilon_d) \equiv G_1/2\sqrt{G_{ref} G_d }$.
In the low temperature limit $T \rightarrow 0$, $r_{G}$ is
equivalent to $r_T(\epsilon_F)$ since $ G_{1/ref/d} =\int
\frac{d\epsilon}{2\pi} T_{1/ref/d}(\epsilon) \frac{-d
f(\epsilon)}{d\epsilon}$. We keep the temperature sufficiently low
($k_B T$ much smaller than the width of the intradot level), so
$r_G$ is very close to $r_T$.

The coherence rate $r_G$, the conductance $G_d$, and the
first-order amplitude $G_1$ versus the QD's level $\epsilon_d$ for
different values of $U$ are shown in Fig.2(c)-(e). $G_0$ or $G_1$
exhibits a peak at $\epsilon_d= 0$ regardless of the value of $U$.
When $U=0$, this peak is symmetric, but at $U\not= 0$, it is
slightly asymmetric. Now we focus on the study of coherence rate $r_{G}$.
First, far away from the resonance peak, i.e. in the co-tunneling
regime, $r_G$ is almost equal to 1 for both $U=0$ and
$U\not=0$. This result implies that the traversing electron keeps
coherence in this regime. Second, in the proximity of the
peak, i.e. in the resonant tunneling regime, $r_G>1$.
This shows that the
higher-order reflecting and tunneling processes still exist and are
not negligible in the resonance regime despite of the open AB
setup with large rate value of $\Gamma/\Gamma^{sd}=5$. Under the
circumstance, $r_T$ (or $r_G$) can not be used to
describe the degree of coherence as
mentioned before.

Therefore the degree of coherence in the resonant regime has to be
further studied. In the following we investigate this question by
using two methods. (i) With increasing $\Gamma /\Gamma^{sd}$, the
bypass effect is enhanced. As $\Gamma/\Gamma^{sd} \rightarrow
\infty$, the higher-order processes completely disappear, left
with only the first-order process. In this limit, we find that
$r_T \rightarrow 1$, regardless of $U$ and $\epsilon_d$ (see
Fig.3a,b). (ii) At finite value of $\Gamma/\Gamma^{sd}$, although
the higher-order and the first-order processes all exist, we can
distinguish them in our calculation, because that the first-order
process has the factor $\Gamma^s\Gamma^d$, while the higher-order
processes has the factors $(\Gamma^s)^2\Gamma^d$,
$\Gamma^s(\Gamma^d)^2$, and so on. So we can withdraw the part of
the contribution of the first-order process in the transmission
probability (or the conductance). And they are: $T_1 =
2|\Gamma^s_{12} \Gamma^d_{34} \tilde{G}^r_{41} \tilde{G}^a_{23}|$,
$T_{ref}=\Gamma^s_{22}\Gamma^d_{33} |\tilde{G}^r_{32}|^2$, and
$T_d=\Gamma^s_{11}\Gamma^d_{44} |\tilde{G}^r_{41}|^2$, where
$\tilde{\bf G}$ is the Green's function when
$\Gamma^s_{mn}=\Gamma^d_{mn}=0$. Following the coherent rate
parameter of only the contribution of the first-order process,
$r_T = T_1/2\sqrt{T_{ref}T_d}$ is exact $1$ regardless of the
values for $U$, $\epsilon_d$ and $\Gamma$. Any one of the results
of (i) and (ii) clearly demonstrates that the electron transport
through the QD is fully coherent, and intradot e-e interaction
does not induce any incoherent effect! In order to check the
reliability of the conclusion reached above, we have purposely
introduced a dephasing source (e.g. virtual Buttiker's voltage
lead\cite{n11}) to the QD, we find that $r_T$ for both (i) and
(ii) are indeed less than $1$ due to the dephasing effect.

Next, we turn to investigate the crossover of
$r_G$ in going from the open multi-terminal AB
interferometer to the closed two-terminal setup (see Fig.3a,b). As
$\Gamma /\Gamma^{sd} \rightarrow \infty$, the setup is completely open and
$r_T=1$. With decreasing of $\Gamma /\Gamma^{sd}$, a peak
emerges at $\epsilon_d=0$. This peak goes up initially and goes down
as $\Gamma /\Gamma^{sd}$ is further reduced. Eventually a valley
emerges at $\epsilon_d =0$, and
the bottom of the valley reaches to zero as
$\Gamma/\Gamma^{sd} \rightarrow 0$ (i.e. a completely closed AB setup). We
emphasize that  $r_G <1$ (or even $r_G \approx 0$) for a closed two-terminal
setup does not imply the occurrence of incoherence. For
example, for $U=0$, it is well known that the electron transport
through the QD is fully coherent, but $r_G \approx 0 $ still in
the vicinity of $\epsilon_d=0$. This clearly means that the
reduction of $r_G$ is from the existence of the higher-order
reflecting and tunneling processes due to the constraint of the
two-terminal interferometer. For the case of $U=\infty$, $r_G$ has
a similar behavior as for the $U=0$ case. The only difference between
them is that the valley is asymmetric for $U=\infty$ and
symmetric for $U=0$. Similarly, one can also not conclude the
appearance of incoherence at $U\not=0$ from $r_G<1$ or asymmetry in
$r_G$. It only shows that the closed two-terminal AB interferometer is not a
suitable setup to quantitatively study the dephasing effect due to
the constraint structure, just as it is not a suitable geometry to
study the phase of the transmission amplitude.\cite{n1,n2,n9,n10}

The same conclusion is reached by studying the amplitude $G_1$ in
the two-terminal AB system. Fig.3c and 3d show our results of
$G_1$ versus $\epsilon_d$ for $U=0$ and $U=\infty$. $G_1$ exhibits
two peaks around $\epsilon_d=0$. Those two peaks are symmetric for
$U=0$, and asymmetric for $U =\infty$ (or a finite $U$). Those
behavior are consistent with the previous theoretical and
experimental findings.\cite{n5,n8} However, we emphasize again
that those results (including the asymmetric peaks at $U\neq 0$)
are the consequences of the confined structures.

Finally, let us turn to the case of applying a magnetic field $B$ to
the QD. Now $\epsilon_{d\uparrow} \neq \epsilon_{d\downarrow}$ due
to the Zeeman splitting. Fig.2f shows the results of $G_d$, $G_1$,
and $r_G$ for $U=\infty$ at $\Delta\epsilon_d
=\epsilon_{d\downarrow} - \epsilon_{d\uparrow} =4$. Both $G_d$ and
$G_1$ have similar results with the zero magnetic field case,
except that the peaks are slightly suppressed. However, $r_G$
is obviously smaller than 1 (see Fig.2f). Even in the limit
$\Gamma/\Gamma^{sd} \rightarrow \infty$, $r_G$ still is less than $1$. Does
this mean that the electron transport through the QD is partially
dephased by $U$ and $B$? To address this question,
we first consider the simple model with $U=0$ and
$\Gamma/\Gamma^{sd}\rightarrow \infty$. While $U=0$, the electron
transport through the QD is fully coherent, and let
$t_{d\uparrow}$ and $t_{d\downarrow}$ describe the transmission
amplitudes for up and down spins. To assume that the transport
through the reference arm is spin independent, then the coherence
rate $r_T$ reduces to:
\begin{eqnarray}
r_T \equiv \frac{T_1}{2\sqrt{T_d T_{ref}}} = \frac{|t_{d\uparrow}
+t_{d\downarrow}|} {\sqrt{2(|t_{d\uparrow}|^2+
|t_{d\downarrow}|^2) }}.
\end{eqnarray}
While $B\not=0$, $t_{d\uparrow}$ is generally not equal to
$t_{d\downarrow}$, then Eq.(4) clearly shows that $r_T <1$
despite of fully coherent transport. For the same reason, $r_G <1$
as shown in Fig.2f for $U=\infty$ is similar as the above case of
$U=0$. Thus, one can not judge incoherence from $r_G<1$ when
$B\not=0$. One has to further study the spin-resolved coherence
rates $r_{G\uparrow}$ and $r_{G\downarrow}$ ($r_{G\sigma} \equiv
\frac{G_{1\sigma}}{2\sqrt{G_{ref\sigma} G_{d\sigma} }}$) for
spin-up and spin-down components (see Fig.2f). Both of them are
very close to $1$. In particularly, in the limit of
$\Gamma/\Gamma^{sd} \rightarrow \infty$, both $r_{T\uparrow}$ and
$r_{T\downarrow}$ approach to $1$ independent of $U$, $B$, and
$\epsilon_{d\sigma}$. Therefore, we conclude that the electron
transport through the QD is fully coherent in the presence of both
$B$ and $U$.

In conclusion, by using an open multi-terminal AB interferometer
we investigate the degree of coherence of the electron transport
through an interacting QD. We demonstrate that the intradot e-e
interaction does not induce any dephasing effects. Furthermore, we
clarify that the asymmetric amplitude in the AB oscillation of the
linear conductance in the two-terminal AB setup originates from
the constraint of this closed setup, and does not reflect
partial dephasing.

{\bf Acknowledgments:} We gratefully acknowledge financial support
from the Chinese Academy of Sciences and NSFC under Grant No.
90303016. XCX is supported by US-DOE.

$^{\ast }$Electronic mail: sunqf@aphy.iphy.ac.cn

\newpage
\begin{figure}

\caption{Schematic diagram for an open multi-terminal AB
interferometer, containing a QD in its lower arm, penetrated by a
magnetic flux $\Phi$, and four extra leads coupled. }\label{fig:1}

\caption{ (a) The total conductance $G$ vs. the intradot level
$\epsilon_d$ at $U=5$, and (b) the flux-dependent conductance
$\Delta G \equiv G(\phi)-G(\phi=0)$ of the three points specified
in (a). (c-f) $G_d$ (dashed), $G_1$ (solid), and $r_G$ vs.
$\epsilon_d$ for the case of $U$=$0$ and $\Delta \epsilon_d=0$
(c), $U=5$ and $\Delta \epsilon_d=0$ (d), $U=\infty$ and $\Delta
\epsilon_d=0$ (e),  and $U=\infty$ and $\Delta \epsilon_d=4$ (f).
In (f), $r_{G\downarrow}$ and $r_{G\uparrow}$ are also plotted
with $\epsilon_{d\uparrow} =\epsilon_d$ and
$\epsilon_{d\downarrow} =\epsilon_d +\Delta\epsilon_d$. The other
parameters in (a-f) are $\Gamma/\Gamma^{sd}=5$ and $t_1=t_4=1$. }
\label{fig:2}

\caption{(a,b) The coherence rates $r_G$ vs. $\epsilon_d$ for
different couplings $\Gamma=10$, 50, 2, 0.5, and 0.1 from the up
to down curves, with $U$=$0$ (a) and $\infty$ (b). (c,d) $G_1$ vs.
$\epsilon_d$ at $\Gamma=0.5$, with $U=0$ (c) and $\infty$ (d). The
parameter $t_1=t_4=1$. } \label{fig:3}

\end{figure}


\begin{references}
\bibitem{addnote}
Here each site $\beta$ only has one level $\epsilon_{\beta\sigma}$.
If there are many levels on each site, our conclusion
remains.
\bibitem{note2}
Here we choose the four sites' levels to be different. In fact, the
results for sufficiently open AB setup is independent of those
levels.
\end{references}
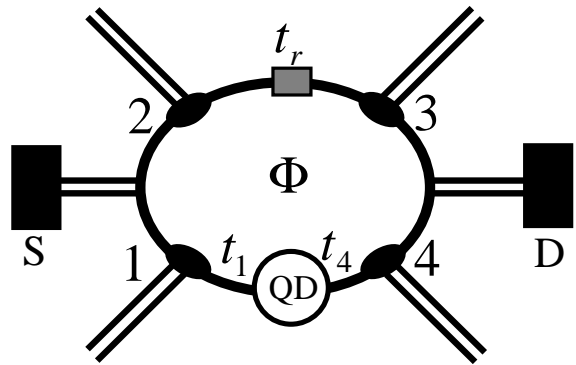
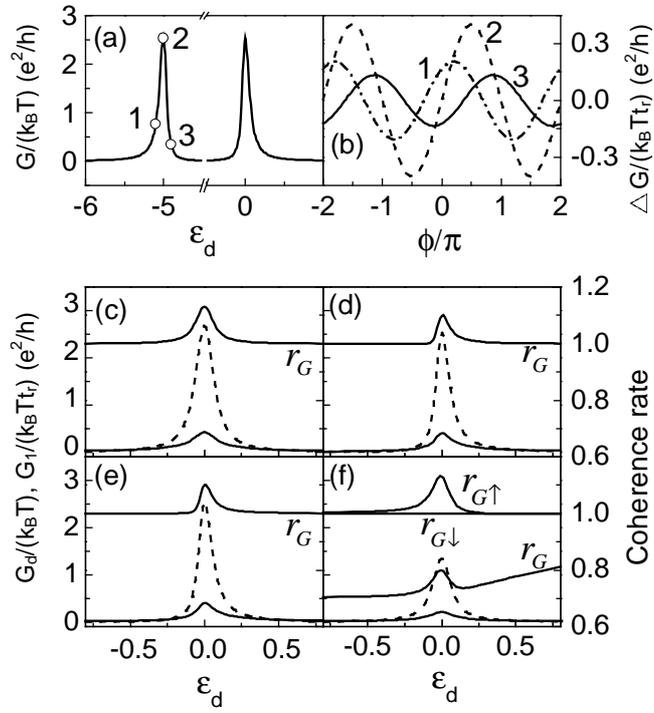
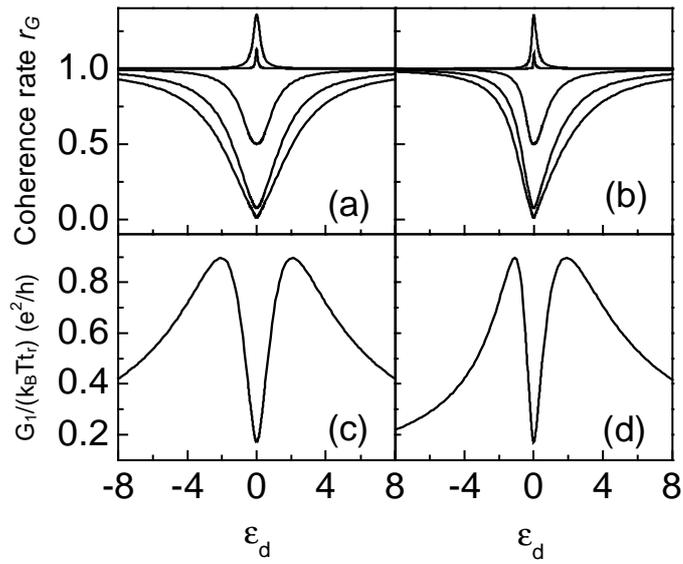
\end{document}